\documentclass[]{interact}

\usepackage{epstopdf}% To incorporate .eps illustrations using PDFLaTeX, etc.
\usepackage{subfigure}% Support for small, `sub' figures and tables
\usepackage{rotating}
\usepackage{xcolor}
\usepackage{float}
\usepackage{geometry}
\usepackage{amssymb}
\usepackage{listings}
\usepackage{amsmath}
\usepackage{pdflscape}
\usepackage{adjustbox}
\usepackage{rotating,booktabs,graphicx,bbm}
\theoremstyle{plain}% Theorem-like structures provided by amsthm.sty

\theoremstyle{definition}

\theoremstyle{remark}

\numberwithin{equation}{section}
\begin{document}

\title{On an Inflated Unit-Lindley Distribution}

\author{
\name{Sudeep R. Bapat\textsuperscript{a} and Rohit Bhardwaj\textsuperscript{b}}
\affil{\textsuperscript{a}Department of Operations Management and Quantitative Techniques, Indian Institute of \\ \hspace{.08 cm} Management, Indore, India\\ \textsuperscript{b}Department of Decision Sciences, Indian Institute of Management, Bangalore, India}
}

\maketitle

\begin{abstract}
Modeling fractional data in various real life scenarios is a challenging task. This paper considers situations where fractional data is observed on the interval [0,1]. The unit-Lindley distribution has been discussed in the literature where its support lies between 0 and 1. In this paper, we focus on an inflated variant of the unit-Lindley distribution, where the inflation occurs at both 0 and 1. Various properties of the inflated unit-Lindley distribution are discussed and examined, including point estimation based on the maximum likelihood method and interval estimation. Finally, extensive Monte Carlo simulation and real-data analyses are carried out to compare the fit of our proposed distribution along with some of the existing ones such as the inflated beta and the inflated Kumaraswamy distributions. 
\end{abstract}


\section{Introduction}\label{UL}

Many a times we come across continuous data which lies strictly between $(0,1)$ and practitioners need to model these using appropriate distributions. Analyzing proportions or percentages etc. fall in this category. One of the widely used distributions is the Beta distribution, which is being applied to such datasets because of its flexibility. However there is a drawback in using the Beta distribution as it is not very appropriate under some real life situations, e.g. hydrological data. Keeping this in the background, [Kumaraswamy, 1976] introduced another similar distribution to counter the Beta model, which is called as the Kumaraswamy distribution. An advantage here is that its distribution and quantile functions can be expressed in closed forms. On the other hand, a Lindley distribution with support on $(0, \infty)$, which was introduced by [Ghitany et al., 2008] is one of the widely discussed distributions over the last decade. One may additionally refer to [Nadarajah et al., 2011] or [Shanker and Mishra, 2013] among others, for extensions and variations of a general Lindley model. Very recently, [Mazucheli et al., 2019a] and [Mazucheli et al., 2020a] introduced the unit-Lindley and the new-unit Lindley distributions having supports on $(0,1)$ and $(0,1]$ respectively, by using the transformations $Y=X/(1+X)$ and $Y=1/(1+X)$ respectively, where $X$ is a suitable Lindley distributed random variable. Some of the key advantages in using the above two models instead of a Beta model are existence of closed forms of the distribution function and the moments. Further, these two models only contain a single parameter and hence are simpler to handle. Other proposed unit-distributions include unit-Gamma [Grassia, 1977], unit-inverse Gaussian [Ghitany et al., 2019], unit-Weibull [Mazucheli et al., 2020b] etc. among many others. 

On a separate note, there are abundant scenarios where such unit datasets also contain $0$'s, $1$'s or both $0$'s and $1$'s. Hence, one needs to assume values which lie between $[0,1)$, $(0,1]$ or $[0,1]$, and analyze such datasets using appropriate inflated models. Such inflated models contain a mixture distribution with the continuous part being modeled through the actual distribution, whereas a degenerate distribution is fitted on $0$'s and $1$'s. In the same spirit, [Ospina and Ferrari, 2010] introduced the inflated beta distribution, whose density function is given by:
\begin{equation}
beinf(y; \alpha, \gamma , \mu, \phi)=
    \begin{cases}
     \alpha (1-\gamma) & \text{if y = 0}\\
     \alpha \gamma & \text{if y = 1}\\
     (1-\alpha) f(y;\mu, \phi) & \text{if y $\in$ (0,1)},\\
  \end{cases}
\end{equation}
where $\alpha$ is the mixing parameter and $f(y;\mu, \phi)$ is the usual density function of a $Beta(0,1)$ distribution.
As a follow-up, [Cribari and Santos, 2019] introduced an inflated Kumaraswamy distribution which was seen to be better than the inflated beta model under some specific real life situations. Its density function can be written down as follows:
\begin{equation}
zoik(y;\lambda,p,\alpha,\theta)=
    \begin{cases}
     \lambda (1-p) & \text{if y = 0}\\
     \lambda p & \text{if y = 1}\\
     (1-\lambda) g(y;\alpha, \beta) & \text{if y $\in$ (0,1)},\\
  \end{cases}
\end{equation}
where $\lambda$ is the mixing parameter and $g(y;\alpha, \beta)$ is the usual density function of a $Kumaraswamy(0,1)$ distribution. In this paper, we try to extend this front by proposing a zero- and one- inflated unit-Lindley or better called as an inflated unit-Lindley (ULINF) distribution in a hope that it will be advantageous according to the earlier mentioned reasons. As for its continuous part, we will focus on the unit-Lindley distribution introduced by [Mazucheli et al., 2019b] whose cumulative distribution function is given by,
\begin{equation}
    F(y;\theta)=1-\left(1-\frac{\theta y}{(1+\theta)(y-1)}\right) exp\left(-\frac{\theta y}{1-y}\right),\;\;\;\;0<y<1; \;\; \theta>0.
\end{equation}
Its corresponding probability density function is given by, 
\begin{equation}\label{1.4}
    f(y;\theta)=\frac{\theta^2}{1+\theta}(1-y)^{-3} exp\left(-\frac{\theta y}{1-y}\right),\;\;\;\;0<y<1; \;\; \theta>0,
\end{equation}
where $\theta>0$, is the only unknown parameter.

Recently, [Ferreira et al., 2020] has proposed a one-inflated unit-Lindley distribution as a preprint. However our current work explores more on an inflated unit-Lindley distribution (inflated at both $0$ and $1$). The structure of the paper is as follows: In Section 2 we introduce the inflated unit-Lindley distribution along with its properties and behavior of its density function over various parameter values. Section 3 talks about maximum likelihood estimation of the unknown parameters. We present extensive simulation studies, comparisons and analysis of practical real datasets in Sections 4 and 5 respectively. We end with brief conclusions and some limitations in Section 6.

\section{The inflated unit-Lindley distribution}
As a build-up towards proposing an appropriate inflated model, we will present a very brief review of finite mixture models. Finite mixture models are widely used to model the data from a statistical population that consists of two or more sub-populations. 

Let $\textbf{y}$ = $\{y_i$, i = 1, ... , n$\}$ be a random sample of size n from an $m$-component finite mixture with the following density:
\begin{equation} \label{3}
    f(y_i; \textbf{$\theta$}) =  \sum_{j=1}^m \alpha_j f_j(y_i; \textbf{$\theta_j$}),
\end{equation}
where $f_j$($y_i$;$\theta$) represent the probability densities (or masses), \textbf{$\theta$} = ($\alpha_1, \alpha_2,..., \alpha_m, \theta_1, \theta_2,..., \theta_m$) and $\alpha_j$ are non negative quantities such that $\alpha_1 + \alpha_2+...+\alpha_m = 1$, i.e., $0 \leq \alpha_j \leq 1$ for $j = 1, 2,..., m$. For a much detailed exposition of finite mixture models, one may refer to \cite{McLachlan} or \cite{Kaye}.

In our case, we propose a 2-component finite mixture model in which one component captures the inflated zeros and ones while another captures the values in $(0,1)$. We use an appropriate Bernoulli distribution as the first component with PMF $f_1(.)$ and CDF $F_1(.)$ and a unit-Lindley distribution as a second component with PDF $f_2(.)$ and CDF $F_2(.)$. Hence, the probability density function $(PDF)$ of the inflated unit-Lindley distribution $(ULINF)$ can be written, by using (\ref{3}), as a weighted sum of these two distributions, and its functional form is given by,
\begin{equation} \label{eq4}
    ulinf(y;\alpha, p, \theta) = \alpha f_1(y;p) + (1-\alpha)f_2(y;\theta),
\end{equation}

while the functional form of its cumulative distribution function $(CDF)$ is given by,
\begin{equation}
    ULINF(y;\alpha, p, \theta) = \alpha F_1(y;p) + (1-\alpha)F_2(y;\theta),
\end{equation}

where $p$ ($0 \leq p \leq 1$) is a Bernoulli proportion, $\theta (\theta > 0$) is the unit-Lindley parameter and $\alpha$ is the mixture proportion or weight as discussed in (\ref{3}).

\begin{figure}[h]
\begin{center}
\includegraphics[scale=0.9]{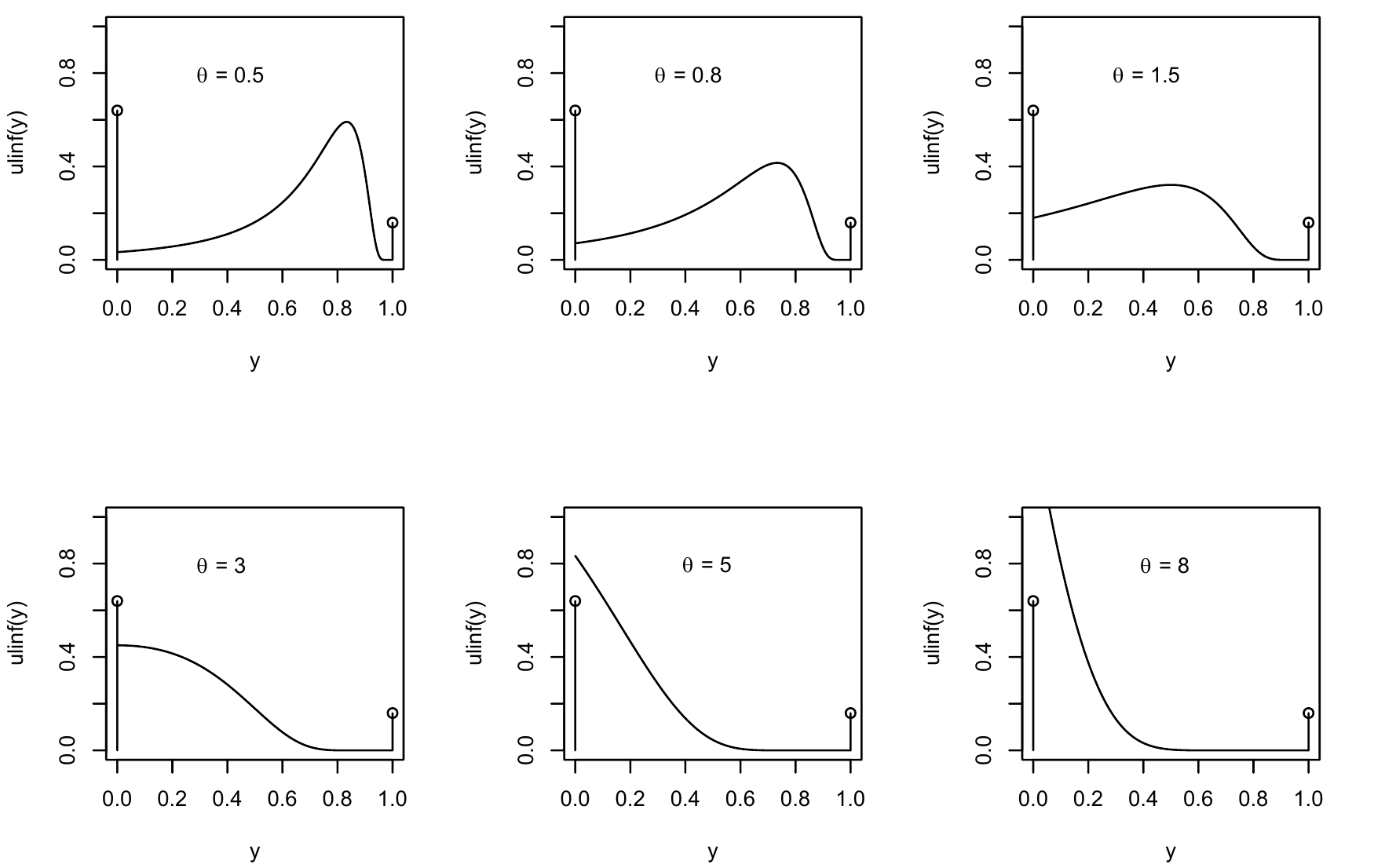}
\caption{\textit{ulinf} densities at different values of $\theta$, with  $p=0.2$ and $\alpha=0.8$}
\label{Figure 1}
\end{center}
\end{figure}

\textbf{Definition 2.1} Let $Y$ be a random variable with support in the interval [0,1]. We say that $Y$ has an inflated unit-Lindley distribution $(ULINF)$ with parameters $\alpha$, $p$ and $\theta$ if its probability density function with respect to the measure space generated by the functional form given in equation (\ref{eq4}) can be written as,

\begin{equation}\label{2.4}
    ulinf(y;\alpha, p, \theta)=
    \begin{cases}
     \alpha (1-p) & \text{if y = 0}\\
     \alpha p & \text{if y = 1}\\
     (1-\alpha) f(y;\theta) & \text{if y $\in$ (0,1)},\\
  \end{cases}
\end{equation}
where $f(y;\theta)$ is the unit-Lindley density as discussed in (\ref{1.4}) with $\theta > 0$ and $0 \leq \alpha, p \leq 1$. We write Y $\sim$ $ULINF(\alpha,p, \theta)$. Note that if Y $\sim$ $ULINF(\alpha,p, \theta)$, then $P(Y=0) = \alpha(1-p)$ and $P(Y=1) = \alpha p$. Further, $P(Y\in(0,1))=(1-\alpha)\int_a^b f(y|\theta)$, for $y \in (0,1)$ and $0 < a < b < 1$. \\

Figure (\ref{Figure 1}) shows the $ULINF$ density plots for different combinations of $\theta$ with fixed $p$ = 0.2 and $\alpha$ = 0.8. Clearly, the density function takes different shapes for different values of $\theta$ and is hence significant where one observes inflated values at the end points of the unit interval.

Further, the $r^{th}$ moment about origin of the $ULINF$ distribution can be written as
\begin{equation} \label{7}
    E(y^r)=\alpha p + (1-\alpha)\mu_r,  \;\;\ r = 1,2,3,...
\end{equation}
where $\mu_r$ is the $r^{th}$ moment about origin of the unit-Lindley distribution. Note that E($y^r$) is the weighted average of the $r^{th}$ moment of a Bernoulli distribution with parameter $p$ and the corresponding $r^{th}$ moment of the $UL(\theta)$ distribution with weights $\alpha$ and $(1-\alpha)$ respectively.
 
Clearly, using (\ref{7}) the mean and variance of Y become:
\begin{equation*}
    E(y) = \alpha p + (1-\alpha)\mu_1 = \alpha p + \frac{1-\alpha}{1+\theta}
\end{equation*}
\begin{equation*}
\begin{split}
    V(y) &= \alpha p + (1-\alpha)\mu_2 -[\alpha p + (1-\alpha)\mu_1]^2 \\
         &= \alpha p + \frac{(1-\alpha)}{1+\theta}(\theta^2 e^{\theta} E_1(\theta)-\theta+1) - \left[\alpha p + \frac{1-\alpha}{1+\theta}\right]^2\\
         &=\alpha p(1-\alpha p) + \frac{(1-\alpha)}{1+\theta}\left(\theta^2 e^{\theta} E_1(\theta)-\theta+1-\frac{1-\alpha}{1+\theta} - 2 \alpha p \right)
\end{split}
\end{equation*}
where $E_1(z)$ = $\int_{1}^{\infty}z^{-a} e^{-xz} dz$ is the exponential integral function. See [Abramowitz and Stegun, 1974] for more details.

\subsection*{Remark 2.1.} As one may have wondered, similar models can be designed if the inflation is only one-sided i.e. only at $0$ or $1$. Analogous mixture density functions as seen in (2.4) can be written down to accommodate such variations. Appropriate moment derivations and maximum likelihood estimation can also be handled. The preprint by [Ferreira et al., 2020] will hopefully be in this direction.

\section{Estimation}
In this section we present the maximum likelihood estimators of the unknown parameters for an inflated unit-Lindley distribution.

\subsection{The likelihood function}
Consider a random sample $\textbf{y} = (y_1,y_2,...,y_n)'$ from the density function given in (2.4) with parameter space $\Lambda$, where $\Lambda$ is a 3 length vector given by $(\alpha,  p, \theta)'$. The corresponding likelihood function is thus given by,
\begin{equation} \label{3.1}
\begin{split}
    L(\Lambda|\textbf{y}) & = \prod_{i=1}^{n} {ulinf(y_i|\alpha,p,\theta)} \\
            & = L_1(\alpha|\textbf{y}) \times L_2(p|\textbf{y}) \times L_3(\theta|\textbf{y})\\
\end{split}
\end{equation}
where
\begin{equation} \label{3.2}
\begin{split}
        L_1(\alpha|\textbf{y}) & = \prod_{i=1}^{n} \alpha^{\mathbbm{1}_{\{0,1\}} (y_i)}    (1-\alpha)^{1-\mathbbm{1}_{\{0,1\}} (y_i)} \\
            & = \alpha^{\sum_{i=1}^n \mathbbm{1}_{\{0,1\}} (y_i)}(1-\alpha)^{n-\sum_{i=1}^n \mathbbm{1}_{\{0,1\}}(y_i)}\\
\end{split}
\end{equation}

\begin{equation} \label{3.3}
\begin{split}
        L_2(p|\textbf{y}) & = \prod_{i=1}^n \left[p^{y_i}(1-p)^{1-y_i}\right]^{\mathbbm{1}_{\{0,1\}} (y_i)} \\
        & = p^{\sum_{i=1}^n y_i \mathbbm{1}_{\{0,1\}} (y_i)}(1-p)^{\sum_{i=1}^n (1-y_i)\mathbbm{1}_{\{0,1\}} y_i} \\
                & = p^{\sum_{i=1}^n \mathbbm{1}_{\{1\}} (y_i)}(1-p)^{\left[\sum_{i=1}^n \mathbbm{1}_{\{0,1\}} (y_i) - \sum_{i=1}^n\mathbbm{1}_{\{1\}} (y_i)\right]}  
\end{split} 
\end{equation} 

\begin{equation} \label{3.4}
\begin{split}
    L_3(\theta|\textbf{y}) & = \prod_{i=1}^{n} f(y_i|\theta) \;\;\;\; for \;\; y_i \in (0,1) \\
    & = \prod_{i=1}^{n} \frac{\theta^2}{1+\theta}(1-y_i)^{-3} exp\left(-\frac{\theta y_i}{1-y_i}\right)\\
\end{split}
\end{equation}

From equation (\ref{3.1}), (\ref{3.2}), (\ref{3.3}),and (\ref{3.4}) we can write the log-likelihood function for $\Lambda = (\alpha, p, \theta)'$ as 

\begin{equation*}
\ell(\Lambda|\textbf{y}) = \ell_1(\alpha|\textbf{y}) + \ell_2(p|\textbf{y}) +  \ell_3(\theta|\textbf{y}),    
\end{equation*}
where
\begin{equation} \label{12}
     \ell_1(\alpha|\textbf{y}) = \ln(\alpha) \sum_{i=1}^n \mathbbm{1}_{\{0,1\}} (y_i) + \ln(1-\alpha) \left[n-\sum_{i=1}^n \mathbbm{1}_{\{0,1\}}(y_i)\right]
\end{equation}

\begin{equation} \label{13}
     \ell_2(p|\textbf{y}) = \ln(p) \sum_{i=1}^n \mathbbm{1}_{\{1\}} (y_i) + \ln(1-p) \left[\sum_{i=1}^n \mathbbm{1}_{\{0,1\}} (y_i)-\sum_{i=1}^n \mathbbm{1}_{\{1\}}(y_i)\right]
\end{equation}

\begin{equation} \label{14}
     \ell_3(\theta|\textbf{y}) = 2n \ln \theta - n \ln(1+ \theta) - \theta t(\textbf{y}) - 3 \sum_{i=1}^n \ln(1-y_i),
\end{equation}
where $t(\textbf{y})$ = $\sum_{i=1}^n\frac{y_i}{1-y_i}$ and $\mathbbm{1}_S(\textbf{y})$ is an indicator variable that takes a value 1 when y $\in S$ and 0 when y $\notin S$.

\subsection{Maximum likelihood estimators}
In order to find out the maximum likelihood estimators, one has to differentiate (\ref{12}), (\ref{13}), (\ref{14}) with respect to the unknown parameters, and equate them to $0$. Following are the first derivatives:

\begin{equation} \label{15}
    \frac{\partial \ell_1(\alpha|y)}{\partial \alpha} = \frac{1}{\alpha} \sum_{i=1}^n \mathbbm{1}_{\{0,1\}} (y_i) - \frac{1}{1-\alpha} \left[n-\sum_{i=1}^n \mathbbm{1}_{\{0,1\}}(y_i)\right]
\end{equation}
\begin{equation} \label{16}
    \frac{\partial \ell_2(p|y)}{\partial p} = \frac{1}{p} \sum_{i=1}^n \mathbbm{1}_{\{1\}} (y_i) - \frac{1}{1-p} \left[\sum_{i=1}^n \mathbbm{1}_{\{0,1\}} (y_i)-\sum_{i=1}^n \mathbbm{1}_{\{1\}}(y_i)\right]
\end{equation}
\begin{equation} \label{17}
    \frac{\partial \ell_3(\theta|y)}{\partial \theta} = \frac{2n}{\theta} - \frac{n}{1+\theta} - t(\textbf{y})
\end{equation}

One can easily see that the maximum likelihood estimators of $\alpha$ and $p$ are hence, $\hat{\alpha}$ = $T_1$/n and $\hat{p}$ = $T_2/T_1$ (0/0 being regarded as 0) where $T_1 = \sum_{i=1}^n \mathbbm{1}_{\{0,1\}} (y_i)$ and $T_2 = \sum_{i=1}^n\mathbbm{1}_{\{1\}} (y_i)$. Here, $\hat\alpha$ is the probability of zeros and ones in the observed sample and $\hat{p}$ is the probability of zeros among the observations that equal zero or one. Since $\hat\alpha$ is an unbiased estimator of $\alpha$ and is a function of the complete sufficient statistic, it is also the UMVUE of $\alpha$ 
\cite[Corollary 1.6.16 and Theorem 1.6.22)]{Casella} and its variance can be written as Var($\hat\alpha$) = $\alpha(1-\alpha$)/n. On the other hand, the maximum likelihood estimator for $\theta$ can be obtained by equating (\ref{17}) to zero and is given by,

\begin{equation}
    \frac{1}{2t(\textbf{y})}\left[n-t(\textbf{y})+\sqrt{t(\textbf{y})^2+6nt(\textbf{y})+n^2}\right]
\end{equation}

The Fisher information matrix for the parameters of the inflated unit-Lindley distribution can be written as 
\begin{equation}\label{3.12}
K(\Lambda) = 
\begin{pmatrix}
k_{\alpha\alpha} & 0 & 0\\
0 & k_{pp} & 0\\
0 & 0 & k_{\theta\theta}\\
\end{pmatrix}
\end{equation}
where
\[
k_{\alpha\alpha} = \frac{n}{\alpha(1-\alpha)}, \;\;\;\;\;  k_{pp} = \frac{n\alpha}{p(1-p)}, \;\;\;\;\;  k_{\theta\theta} = \frac{2n}{\theta^2} - \frac{n}{(1+\theta)^2} \]

Here, $\alpha$, p, and $\theta$ are orthogonal parameters and hence the respective components of the score vector are uncorrelated.

\textbf{Proposition 3.1.} The inflated unit-Lindley distribution belongs to the exponential family of distributions having full rank. Hence, for large samples, it follows that 

\begin{equation*}
    \sqrt{n}(\hat{\Lambda}-\Lambda) \overset{d}\longrightarrow N_3(0,K(\Lambda)^{-1}),
\end{equation*}

where $\Lambda = (\alpha, p, \theta)'$ and $K(\Lambda)$ is a 3 $\times$ 3 fisher information matrix which is as given in (\ref{3.12}). Now, let m($\Lambda$) be a continuous and differentiable function of $\Lambda$. Then, the asymptotic distribution of m($\hat{\Lambda}$), which is the ML estimator of m($\Lambda$), can be obtained using the delta method \cite[Sec. 1.9]{Casella}. We hence get,

\begin{equation*}
    \sqrt{n}(m(\hat{\Lambda})-m(\Lambda)) \overset{d}\longrightarrow N_3(0, b(\Lambda))
\end{equation*}
\noindent
where $b(\Lambda) = \dot{m}(\Lambda)^{T}K(\Lambda)^{-1}\dot{m}(\Lambda)$ with $\dot{m}(\Lambda) = \partial m(\Lambda)/\partial \Lambda$.

Following this, one can construct approximate confidence intervals based on the asymptotic normality $\hat{\Lambda}$. Using the limiting distribution, we can construct the confidence intervals for $\alpha$, p, and $\theta$. Hence, (1-$\alpha^*$) $\times$ 100\% asymptotic confidence intervals for $\alpha$, p, and $\theta$ are given by $\hat{\alpha}\pm z_{(1-\alpha^*/2)}$ se$(\hat{\alpha})$, $\hat{p}\pm z_{(1-\alpha^*/2)}$ se$(\hat{p})$, and $\hat{\theta}\pm z_{(1-\alpha^*/2)}$ se$(\hat{\theta})$ respectively, where $\alpha^*$ is the required significance level.

\section{Analysis from simulations}
In this section, we carry out extensive simulations to gauge the behaviour of the parameters underlying our proposed inflated unit-Lindley distribution. We focus on point estimation, and Tables 1 and 2 contain the relative biases and mean squared errors of the maximum likelihood estimators of $\theta$ and $p$, and also of the mean $(\hat{\mu}_y)$ and variances $(\hat{\sigma}^2_y)$ of the observed sample data. We fix the values of $\alpha=0.25, \theta=1.5$ and $p=0.4$ in Table 1, whereas $\alpha=0.5, \theta=1.5$ and $p=0.4$ in Table 2. We report results for varying sample sizes from small (50) to large (1000) and run 10,000 simulations. All the analysis was carried out in \textsf{RStudio} [RStudio Team, 2020] and used the \texttt{LindleyR} package developed by [Mazucheli et al., 2019a]. The \texttt{R} Code for the simulation study can be found in Appendix A.

In both the tables, the mean squared error is computed as the average squared difference between the estimated value and the true value of the parameter and one can clearly see that the MSE decreases as the sample size increases. We also observed that the estimates are converging at a slow rate as we increase the sample size from 50 to 1000. The convergence for the estimate of p approaches the true value from below whereas for $\theta$, it approaches to the true value from above. For e.g., for $n=50$, the estimate of p and $\theta$ equal 0.3968 and 1.5292 respectively, whereas for $n=1000$, the estimates are 0.3996 and 1.5010 respectively.

\begin{figure}[H]
\centering
\includegraphics[scale=1]{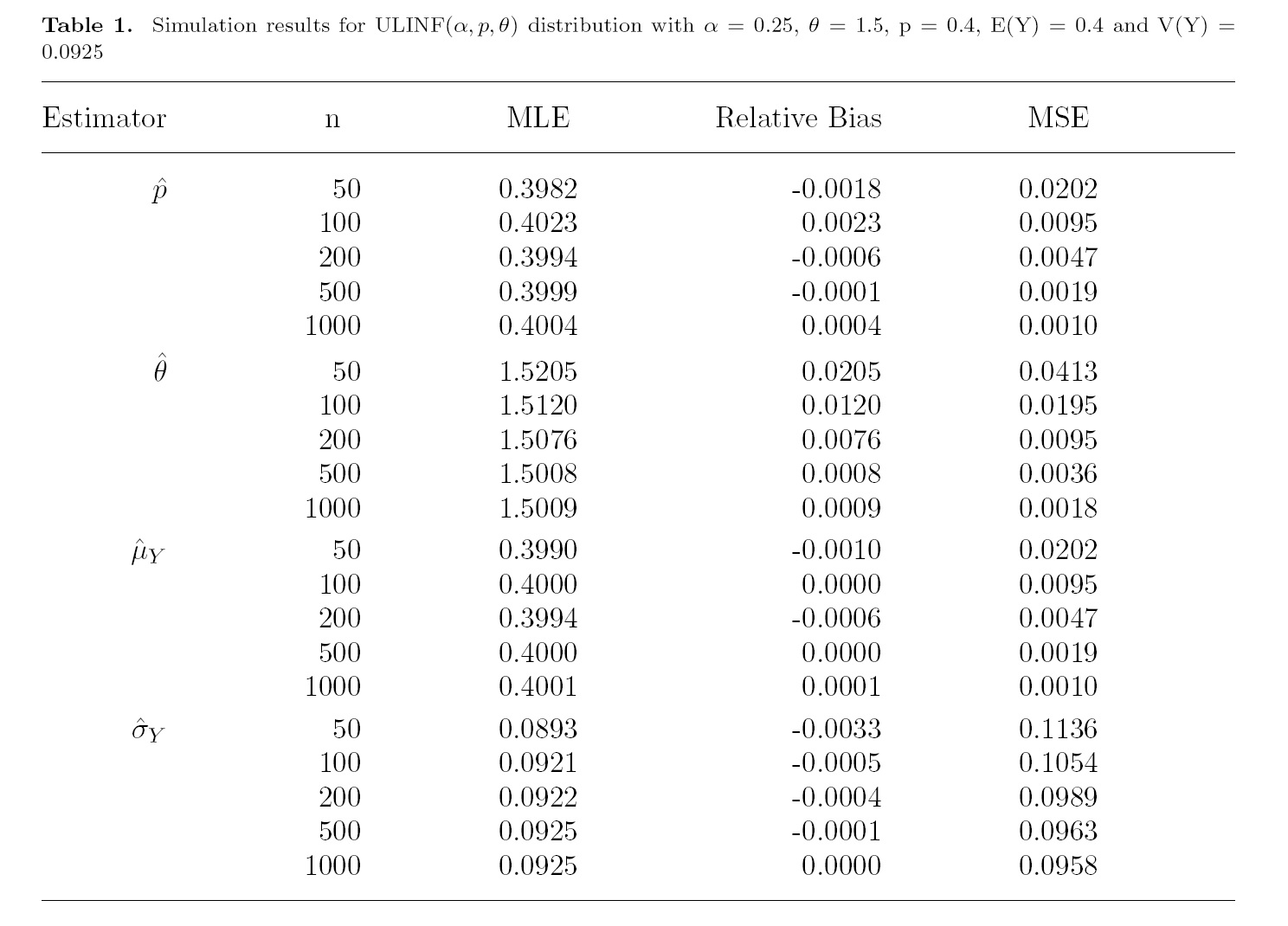}
\end{figure}

\begin{figure}[H]
\centering
\includegraphics[scale=1]{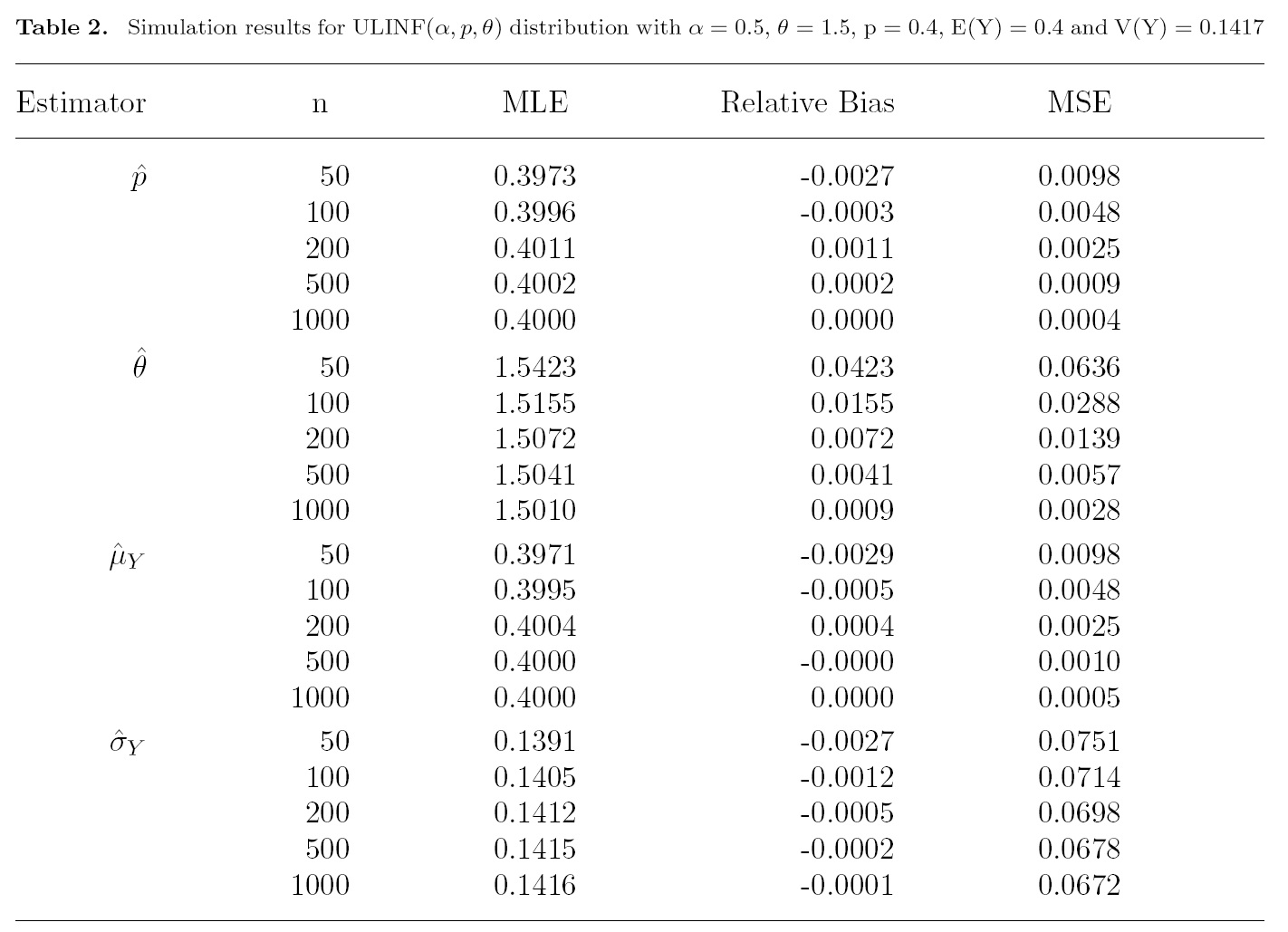}
\end{figure}

\section{Analysis from real data}
In this section, we present real life applications of our proposed distribution to illustrate its potentiality under different types of data sets. We compare our proposed model with the inflated beta and the inflated Kumaraswamy distributions and use AIC and BIC as model selection criteria. Computations for fitting the inflated beta and the inflated Kumaraswamy models were done by using the \texttt{Rfast}
and \texttt{Rfast2} packages, proposed by [Papadakis et al., 2019] and [Papadakis et al., 2020] respectively, which can be found in the CRAN packages repository.

\subsection{Data set I}
The first data set contains the percent of newborn elephants who had their heads up at halfway during pregnancy, among different herd sizes. The data is taken from \textcolor{blue}{https://stackoverflow.com/questions/55984490/arcsine-transformation-of-percentage-data}. The data set contains 27 observations with 2 zeros and 6 ones, and is as follows:\\

\noindent
0.0000, 1.0000, 0.8000, 0.2500, 0.5714, 1.0000, 0.0000, 0.2500, 0.5000, 1.0000, 1.0000, 0.7000, 1.0000, 0.1429, 0.2667, 1.0000, 0.5000, 0.4000, 0.6765, 0.4359, 0.0541, 0.4490, 0.4150, 0.6923, 0.1429, 0.0707, 0.0605.\\

\begin{figure}[H]
\centering
\includegraphics[scale=1]{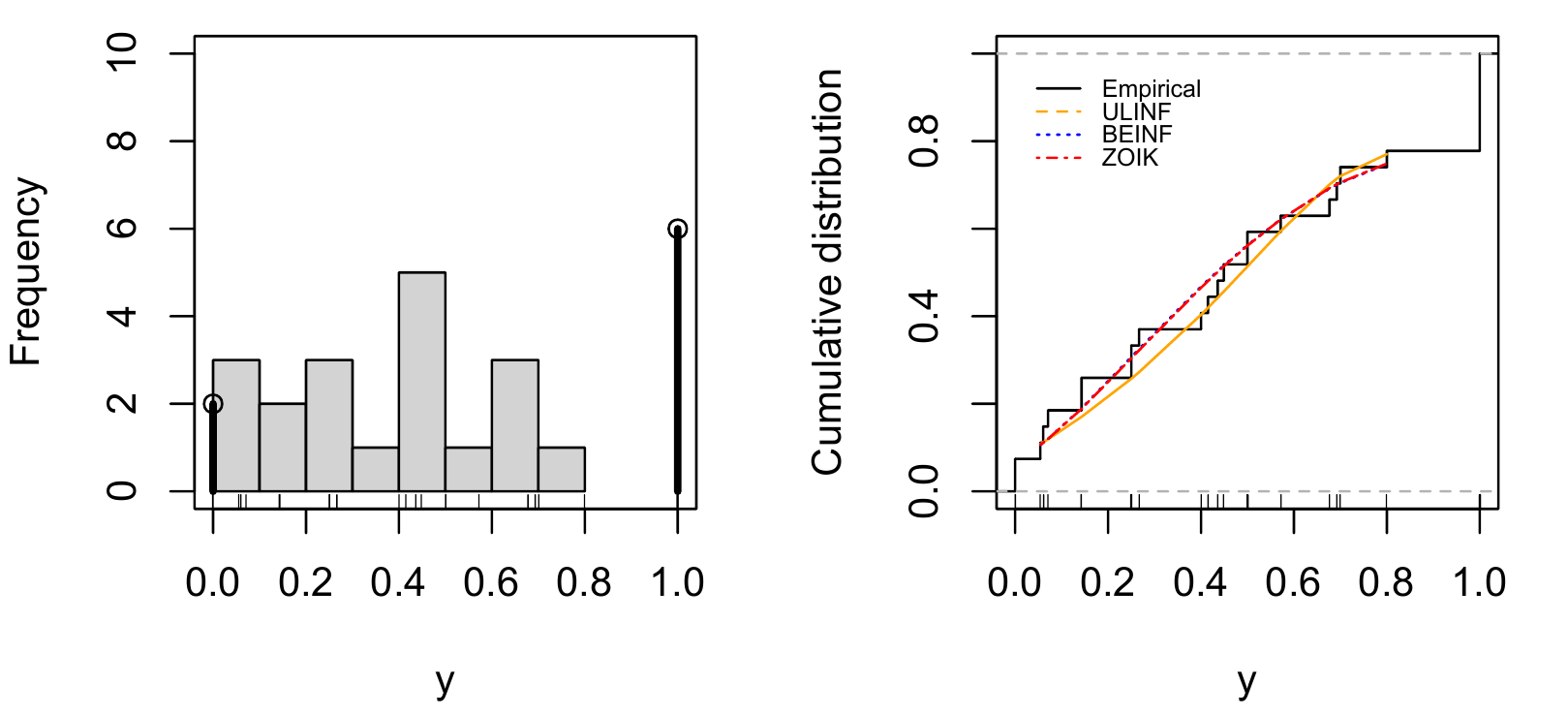}
\caption{Frequency histogram and estimated cumulative distribution functions}
\label{fig2}
\end{figure}

\begin{figure}[H]
\centering
\includegraphics[scale=1]{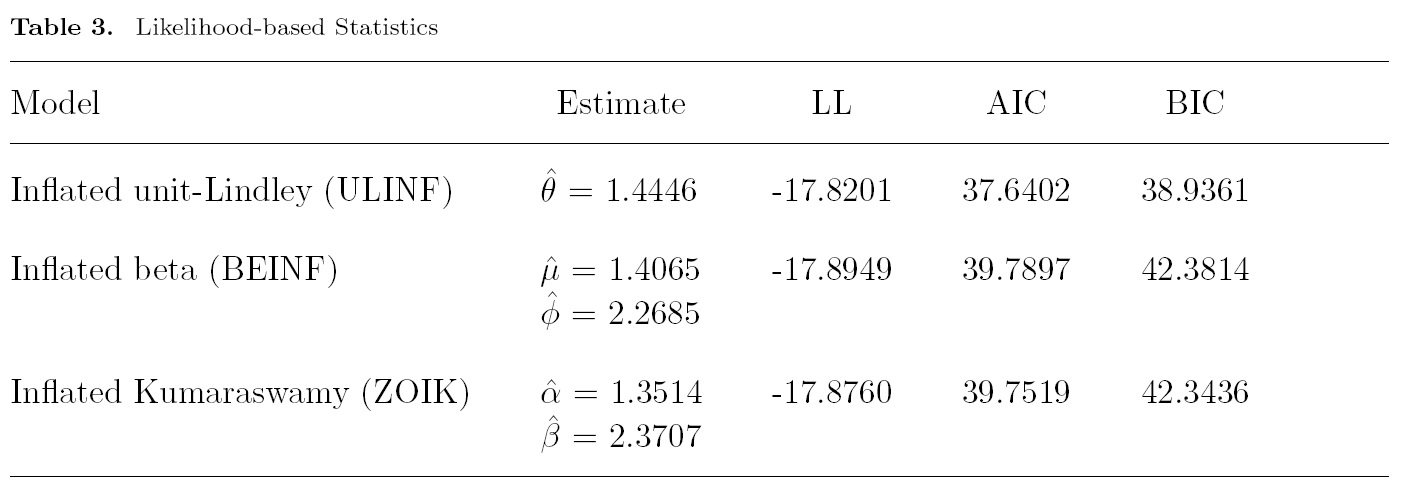}
\end{figure}

For this data, we fit the inflated unit-Lindley distribution, ULINF$(\alpha, p, \theta)$ and compare it with the inflated beta, BEINF$(\alpha, \gamma, \mu, \phi)$ and the inflated Kumaraswamy distribution, ZOIK$(\lambda, p, \alpha, \beta)$. The maximum likelihood estimates of the parameters of the ULINF distribution are $\hat{\alpha}$ = 0.2963, $\hat{p}$ = 0.75 and $\hat{\theta}$ = 1.4446, for the BEINF distribution are $\hat{\alpha}$ = 0.2963, $\hat{\gamma}$ = 0.75, $\hat{\mu}$ = 1.4065, $\hat{\phi}$ = 2.2685 and for the ZOIK distribution are $\hat{\lambda}$ = 0.2963, $\hat{p}$ = 0.75, $\hat{\alpha}$ = 1.3514, $\hat{\beta}$ = 2.3707. Table 3 contains these values along with the log-likelihood (LL), AIC and BIC values for the three distributions.

Figure 2 shows the histogram and the empirical distribution of the data along with the estimated cumulative distribution functions for the inflated unit-Lindley, inflated beta and inflated Kumaraswamy. The plot clearly shows that the inflated unit-Lindley is a better fit for the data as compared to the other two distributions.

\subsection{Data set II}
In spirit of providing a valid illustration having a higher sample size, we generate a pseudo-real data set containing 300 observations with 30 zeros, 50 ones and the remaining 220 observations lying strictly between 0 and 1. Table 4 shows a few summary statistics of the dataset, just for comparison. A suitable R code for generating the hypothetical data can be found in Appendix B.

\begin{figure}[H]
\centering
\includegraphics[scale=1]{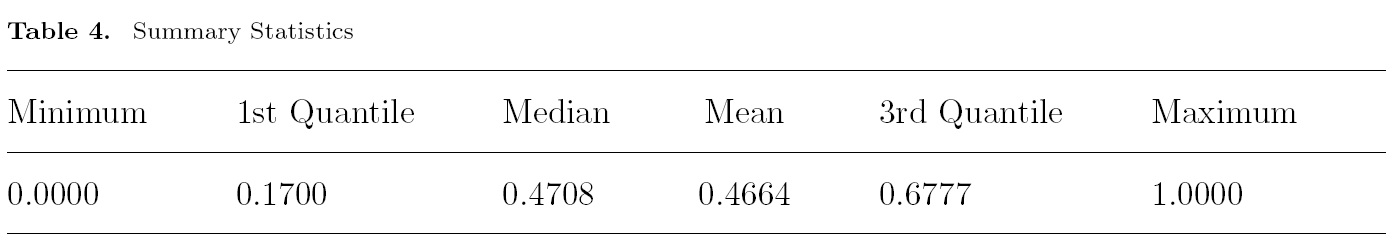}
\end{figure}

The maximum likelihood estimates of the parameters of ULINF, BEINF and ZOIK with their corresponding log-likelihood, AIC, and BIC value for the hypothetical data is shown in Table 5. The ML estimates of the Bernoulli parameter $p$ and the mixing parameters $\alpha, \lambda$ are equal for all the distributions used in the analysis and are $\hat{p}$ = 0.625 and $\hat{\alpha}=\hat{\lambda}$ = 0.2667 respectively. We observe that the AIC and BIC values for the inflated unit-Lindley distribuiton are less than the corresponding values for the inflated beta and the inflated Kumaraswamy distributions. This signifies that our proposed model is better than the other two.

\begin{figure}[H]
\centering
\includegraphics[scale=1]{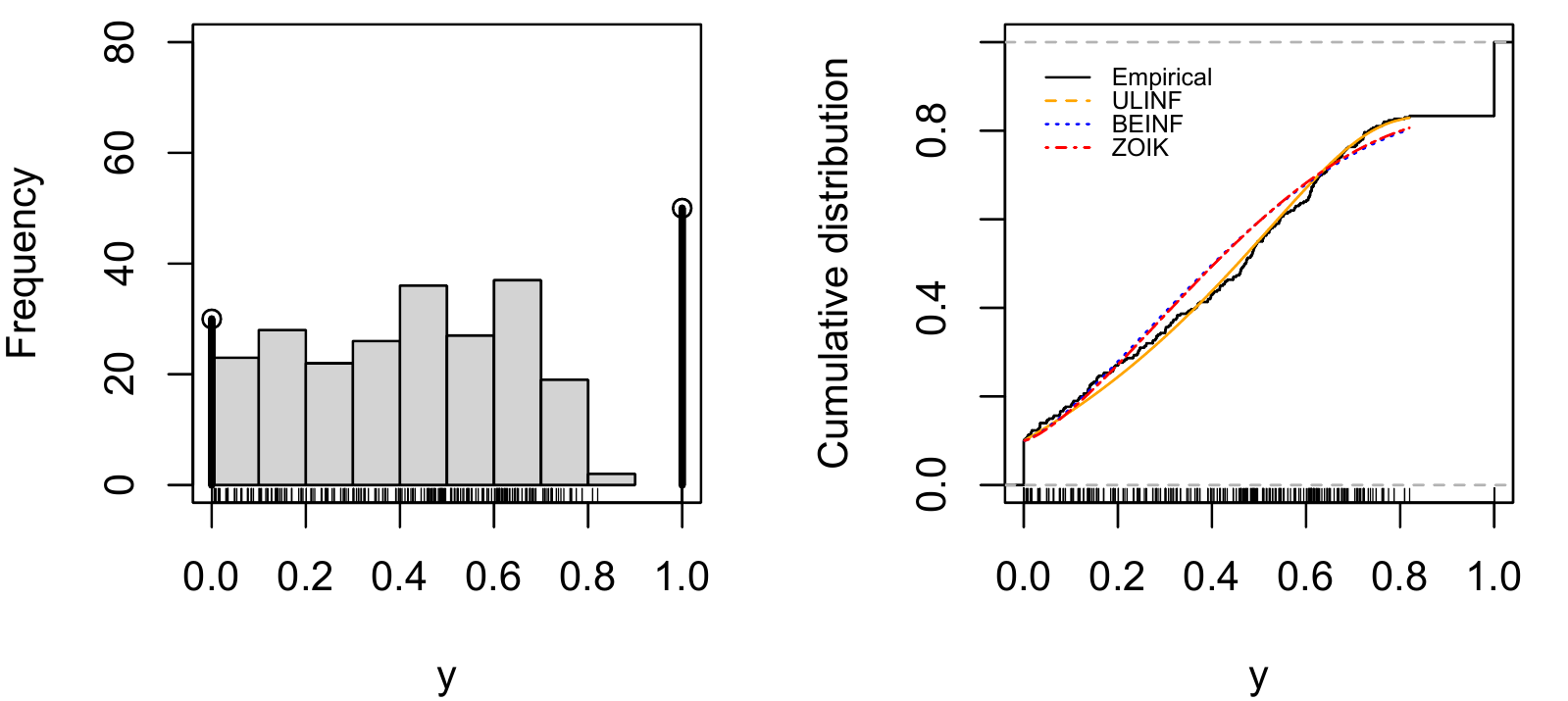}
\caption{Frequency histogram and estimated cumulative distribution functions}
\label{fig3}
\end{figure}

\begin{figure}[H]
\centering
\includegraphics[scale=1]{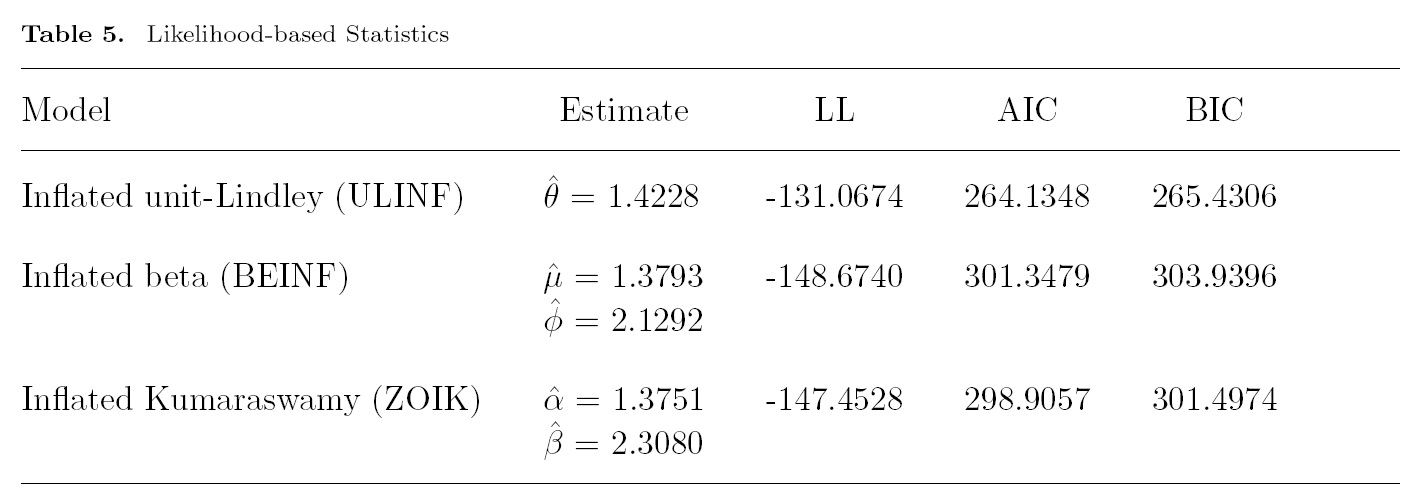}
\end{figure}

Figure 3 depicts a histogram of the dataset and the empirical distribution function along with the fitted distributions namely, ULINF, BEINF and ZOIK for the hypothetical data. Visually, it is evident that among others, the inflated unit-Lindley distribution fits the dataset better.

\section{Limitations and Conclusions}

In this paper, we introduced an inflated version of the unit-Lindley distribution to model datasets which are inflated at both 0 and 1. Appropriate density and distribution functions were derived along with expressions for moments and maximum likelihood estimates for the parameters. Extensive simulation and real data analyses were carried out to compare our proposed model with some of the existing ones such as the inflated beta and the inflated Kumaraswamy distributions, where it was observed that the inflated unit-Lindley performed better than the other two. 

We would like to point out a few limitations in terms of data acquisition for our proposed model. The model is suitable for situations where the parameter $\theta$ of the unit-Lindley distribution takes values more than 1.3 approximately, i.e., there are very less data points in between quantile 3 and quantile 4. Such right skewed unit-Lindley works well considering any values of parameters $\alpha$ and $p$. Observing such kind of data in real life may be a challenging task, but in such case inflated unit-Lindley is the appropriate model to fit.

\section*{References}

Abramowitz, M. and Stegun, I. (1974). \textit{Handbook of Mathematical Functions with Formulas, Graphs and Mathematical Tables}. Dover Publications, Incorporated, New York.\\ \\
Cribari, N. and Santos, J. (2019). Inflated Kumaraswamy distributions. \textit{Annals of the Brazilian Academy of Sciences}, 91(2).\\ \\
Ferreira, A., Mazucheli, J., and Puziol de Oliveira, R. (2020). The one-inflated unit-Lindley distribution. DOI: 10.13140/RG.2.2.29162.34243.\\ \\
Ghitany, M., Atieh, B., and Nadarajah, S. (2008). Lindley distribution and its application. \textit{Mathematics and Computers in Simulation}, 78:493-506.\\ \\
Ghitany, M. E., Mazucheli, J., Menezes, A. F. B., and Alqallaf, F. (2019).
The unit-inverse Gaussian distribution: A new alternative to two-parameter distributions on
the unit interval. \textit{Communications in Statistics - Theory and Methods}, 48(14):3423-3438.\\ \\
Grassia, A. (1977). On a family of distributions with argument between 0 and 1
obtained by trans-formation of the Gamma distribution and derived compound distributions.
\textit{Australian Journal of Statistics}, 19:108-114. \\ \\
Kumaraswamy, P. (1976). Sine power probability density function. \textit{Journal of Hydrology}, 31:181-184.\\ \\
Lehmann, E. and Casella, G. (1998). \textit{Theory of point estimation}. Springer, New York, 2nd ed edition.\\ \\
Mazucheli, J., Bapat, S. R., and Menezes, A. (2020a). A new one-parameter unit-Lindley distribution. \textit{Chilean Journal of Statistics}, 11(1):53-67.\\ \\
Mazucheli, J., Fernandes, L., and Oliveira, R. (2019a). \textit{LindleyR: The Lindley Distribution and its Modifications}. R package version 1.1.0.\\ \\
Mazucheli, J., Menezes, A. F. B., and Chakraborty, S. (2019b). On the one parameter unit Lindley distribution and its associated regression model for proportion data. \textit{Journal of Applied Statistics}, 46(4):700-714.\\ \\
Mazucheli, J., Menezes, A. F. B., Fernandes, L. B., de Oliveira, R. P. and Ghitany, M. E. (2020b). The unit-Weibull distribution as an alternative to the Kumaraswamy distribution for the modeling of quantiles conditional on covariates. \textit{Journal of Applied Statistics}, 47(6):954-974.\\ \\
McLachlan, G. and David, P. (2000). \textit{Finite mixture models}. John Wiley and Sons, New York.\\ \\
McLachlan, G. J. and Kaye, E. B. (1988). \textit{Mixture models, Inference and applications to clustering, volume 1}. Dekker, New York. \\ \\
Nadarajah, S., Bakouch, H., and Tahmasbi, R. (2011). A generalized Lindley distribution. \textit{Sankhya B}, 73.\\ \\
Ospina, R. and Ferrari, S. (2010). Inflated beta distributions. \textit{Statistical Papers}, 51(1):111-126.\\ \\
Papadakis, M., Tsagris, M., Dimitriadis, M., Fafalios, S., Tsamardinos, I., Fasiolo, M., Borboudakis, G., Burkardt, J., Zou, C., Lakiotaki, K. and Chatzipantsiou., C.(2020). Rfast: \textit{A Collection of Efficient and Extremely Fast R Functions}. R package version 2.0.0.\\ \\
Papadakis, M., Tsagris, M., Fafalios, S. and Dimitriadis., M. (2019). \textit{Rfast2: A Collection of Efficient and Extremely Fast R Functions II}. R package version 0.0.5.\\ \\
RStudio Team (2020). \textit{RStudio: Integrated Development Environment for R}. RStudio, PBC., Boston, MA.\\ \\
Shanker, R. and Mishra, A. (2013). A quasi Lindley distribution. \textit{African Journal of Mathematics and Computer Science Research}, 6(4):64-71.

\newpage
\centering{\Large{\textbf{Appendices}}}

\appendix

\raggedright{\section{R function used for simulation study}} \label{Appa}
\begin{lstlisting}[language=R]
rm(list = ls())
library(LindleyR)
library(expint)

f = function(p, alpha, theta){
  
  ruy = numeric()
  t_y = numeric()
  rby = numeric()
  
  b <- c(50, 100, 200, 500, 1000)
  
simstudy <- matrix(0, nrow=15, ncol=5, byrow = TRUE)

rownames(simstudy) <- c("Bias.alpha", "Bias.theta", "Bias.p", "MSE.alpha", 
"MSE.theta", "MSE.p", "alpha.est", "theta.est", "p.est", "E_y", "V_y", "bias.E", "bias.V", "mse.E", "mse.V")

colnames(simstudy) <- c(b[1], b[2], b[3], b[4], b[5])

#True mean and variance 
mean = alpha * p + ((1 - alpha)/(1 + theta))
Variance = alpha * p + ((1 - alpha)/(1 + theta)) * ((theta^2) * expint_Ei(theta) - theta + 1)

for(k in 1:5)
  {
  n <- 2000
  dummy = matrix(5*n, nrow = n, ncol = 5, byrow = TRUE)
  
  for(i in 1:n)
  {
  w <- b[k]
    for(j in 1:w)
    {
      rl <- rlindley(1, theta)
      ruy[j] <- rl/(1 + rl)      #unit-Lindley random variables
      rby[j] <- rbinom(1, 1, p)  #bernoulli random variables
    }
      #partition data using alpha
      uy <- sample(ruy, (1 - alpha) * w, replace = FALSE)
      by <- sample(rby, alpha * w, replace = FALSE)

      y <- sort(c(uy, by))
      x <- ifelse(y == 1, 1, 0)
      T2 <- sum(x)
      
      a <- ifelse(y==0 | y==1, 1, 0)
      T1 <- sum(a)
      
      alpha_hat <- T1/w   # estimate of mixture parameter, alpha
      p_hat <- T2/T1      # estimate of Bernoulli parameter p
  
      len <- length(uy) 
      for(pp in 1: len)
        {
      t_y[pp] <- uy[pp]/(1 - uy[pp])
        }
      ty <- sum(t_y)
      # estimate of unit lindley parameter, theta
      theta_hat <-(1/(2 * ty)) * (len - ty + sqrt(ty^2 + 6 * len * ty + len^2)) 
      
  dummy[i,1] = alpha_hat
  dummy[i,2] = theta_hat
  dummy[i,3] = p_hat
  dummy[i,4] = dummy[i,1] * dummy[i,3] + ((1 - dummy[i,1])/(1 + dummy[i,2]))
  dummy[i,5] = dummy[i,1] * dummy[i,3] + ((1 - dummy[i,1])/(1 + dummy[i,2])) * (dummy[i,2]^2 * expint_Ei(dummy[i,2]) - dummy[i,2] + 1)
  }
  dummy <- dummy[complete.cases(dummy),]
  
  rw <- nrow(dummy)
  
  bias.alpha = mean(dummy[,1] - alpha)
  bias.theta = mean(dummy[,2] - theta)
  bias.p = mean(dummy[,3] - p)
  bias.E = mean(dummy[,4] - mean)
  bias.V = mean(dummy[,5] - Variance)
  
  mse.alpha = (1/rw) * sum((dummy[,1] - alpha)^2)
  mse.theta = (1/rw) * sum((dummy[,2] - theta)^2)
  mse.p = (1/rw) * sum((dummy[,3] - p)^2)
  mse.E = (1/rw) * sum((dummy[,3] - mean)^2)
  mse.V = (1/rw) * sum((dummy[,3] - Variance)^2)
  
  alpha.est = sum(dummy[,1])/rw
  theta.est = sum(dummy[,2])/rw
  p.est = sum(dummy[,3])/rw
  E_p_est = sum(dummy[,4])/rw
  V_p_est = sum(dummy[,5])/rw
  
  simstudy[1,k] <- bias.alpha
  simstudy[2,k] <- bias.theta
  simstudy[3,k] <- bias.p
  
  simstudy[4,k] <- mse.alpha
  simstudy[5,k] <- mse.theta
  simstudy[6,k] <- mse.p
  
  simstudy[7,k] <- alpha.est
  simstudy[8,k] <- theta.est
  simstudy[9,k] <- p.est
  
  simstudy[10,k] <- E_p_est
  simstudy[11,k] <- bias.E
  simstudy[12,k] <- mse.E
  
  simstudy[13,k] <- V_p_est
  simstudy[14,k] <- bias.V
  simstudy[15,k] <- mse.V
}
simstudy
write.csv(simstudy,"simulation_result.csv")
}
\end{lstlisting}

\vspace{5mm}

\section{R code for generating the hypothetical data} \label{AppB}

\begin{lstlisting}[language=R]
rm(list = ls())
library{LindleyR}
set.seed(99)
theta <- 1.444589  
for(p in 1:190)
  {
    h[p] <- rlindley (1, theta)
    hul[p] <- h[p] / (1 + h[p])
  }
hdata <- c(rep(0,20),hul, rep(1,60))   #Hypothetical data
\end{lstlisting}

\end{document}